\documentclass[twocolumn%
              ,showkeys%
              ,jpb]{revtex4-2}

\usepackage{amsmath, amssymb}
\usepackage{graphicx}
\usepackage{xcolor}
\usepackage{bm}

\begin{document}

\title[Sample title]{The role of tunneling in the ionization of atoms by ultrashort and intense laser pulses}

\author{Gabriel M.~Lando}
\email{lando@universite-paris-saclay.fr}
\affiliation{Universit{\'e} Paris-Saclay, CNRS, LPTMS, 91405 Orsay, France}

\date{\today}

\begin{abstract}
    Classically allowed transport is shown to compete with quantum tunneling during the ionization of atoms by ultrashort and intense laser pulses, despite Keldysh parameters smaller than unity. This is done by comparing exact probability densities with the ones obtained from purely classical propagation using the Truncated Wigner Approximation. Not only is classical transport capable of moving trajectories away from the  core, but it can also furnish ionization probabilities of the same order as the quantum ones for intensities currently employed in experiments. Our results have implications ranging from a conceptual correction to semiclassical step models in strong-field physics to the ongoing debate about tunneling time measurements in attoclock experiments.
\end{abstract}

\maketitle

\emph{Introduction --} Tunneling is one of the most characteristically quantum phenomena in nature. Often depicted as a near-magical violation of classical conservation laws, it is the basic mechanism behind several physical and chemical processes and technologies, such as: the stability of star cores \cite{Itoh1979}, the decay of radioactive elements \cite{Gurney1928}, Josephson junctions \cite{Josephson1962}, the transfer of hydrogen atoms in chemical reactions \cite{Kohen1999,*hammes2006hydrogen,*Lowdin1963, *Trixler2013}, photosynthesis \cite{Peters1978}, electron transfer in proteins \cite{gray2003electron} and flash memory cards \cite{esaki1974long,*Bez2003}, and a multitude of others. It is then remarkable that, to this day, the use of the term ``tunneling'' is not uniform among all areas of physics and chemistry. This is especially true when the phenomenon takes place in time domain, as opposed to the static tunneling tails penetrating classically forbidden regions in quantum eigenstates \cite{berry1972semiclassical}.

For conservative systems with a single degree of freedom (DoF), it is tempting to interpret time-dependent tunneling as the transport of portions of the wave function through a tall potential barrier. However, this description does not take momentum into account, and it is well-known that a more accurate picture is given in terms of the Wigner function \cite{balazs1990wigner}. Here, it is possible to directly visualize the full energy barrier in phase space, which forms a separatrix, instead of its misleading projection in position space only. One then sees that the Wigner function naturally extends over the momentum direction and presents high-energy tails \cite{balazs1990wigner,berry1972semiclassical,berry1977semi,ozorio1998weyl}. If points are sampled on these tails and evolved according to Hamilton's equations, they will emerge on the other side of the barrier through classically allowed transport. This contrasts with the case of tunneling, where transport is necessarily classically forbidden \cite{maitra1997barrier,grossmann1995tunneling,grossmann2000tunneling2, wang2021semiclassical}. 

The matter of whether classically allowed or forbidden transport is the dominant mechanism for some particular phenomenon is relatively straightforward for the simple systems in the above paragraph, but becomes a lot more blurred in the high-dimensional or time-dependent case: The former is deceptive because splitting phase space into disjoint regions is harder in higher dimensions \cite{spanner2003strong,zagoya2014interference}; The latter involves non-stationary potential barriers through which energy is added to and/or subtracted from the system, possibly triggering chaos even for a single DoF \cite{chirikov1979universal,ArnoldBook,OzorioBook}. For both cases a proper definition of tunneling is quite challenging, although progress has been recently made in this direction \cite{wang2021semiclassical}. 

The interactions between atoms and ultrashort, intense laser pulses belong to the category of time-dependent processes with strongly chaotic classical dynamics. Here, the mechanism referred to as tunneling ionization (TI) lies at the heart of a plethora of phenomena, forming the first stage of several semiclassical step models \cite{corkum1993plasma,lewenstein1994theory}. There are good reasons to question the role of tunneling in this context, as the overall classical effect of the pulse is a chaotic shaking of classical trajectories, which are then perfectly able to escape the core \emph{via} classically allowed ionization (CAI) \cite{dubois2019thesis}. The number of trajectories scattered by CAI is even enough to reproduce higher-harmonics generation (HHG) spectra semiclassically, despite the propagated state being approximately the atom's ground state \cite{zagoya2014quantum}. Since semiclassical calculations rely completely on classical trajectories, this would be impossible if the transport pathways were truly dominated by TI \cite{maitra1997barrier}. 

In this manuscript, we do not make the standard distinction between vertical (multi-photon) and horizontal (tunneling) ionization channels \cite{GrossmannBook}. Instead, we distinguish CAI from TI by direct comparisons with purely classical simulations, where the distribution of initial trajectories does not come from any tweaking \cite{ni2016tunneling,van1999irregular,hofmann2019attoclock}, but from the system's \emph{true} ground state. Since tunneling is stricter and easier to spot in systems with a single DoF, our simulations are performed using the improved soft-core potential of Majorosi \emph{et al} \cite{majorosi2018improved} coupled to an intense, ultrashort and linearly polarized (LP) laser pulse in the near-infrared range. Contrary to intuition, but in line with \cite{spanner2003strong, zagoya2014quantum}, we demonstrate that CAI and TI are deeply intertwined, and possibly inseparable. Among the implications of these results lie a correction to semiclassical step models, since ionization is mostly unrelated to tunneling, and an added difficulty to the ongoing debate on ``tunneling'' time measurements \cite{landsman2014ultrafast, torlina2015interpreting, ni2016tunneling, pollak2017quantum,rost2019attoclock,hofmann2019attoclock,kheifets2020attoclock,sainadh2020attoclock,hofmann2021quantum}. 

\noindent
\emph{Model system--} One-dimensional systems allow for visualization ease, but often disagree with full 3 DoF simulations -- at least quantitatively. The popular soft-core potential, for example, has been shown to largely overestimate ground-state depletion and other expectation values \cite{majorosi2018improved}. Nevertheless, recent work by Majorosi \emph{et al} has shown that the improved soft-core system (in atomic units)
\begin{equation}
    H_{\rm{iSC}}(p_x,x) = p_x^2/2 - Z \left( Z^{-2} + 4x^2 \right)^{-1/2} \quad  \label{eq:isc}
\end{equation}
offers strikingly accurate estimates when compared to simulations in full dimensionality using the exact Coulomb potential \cite{majorosi2018improved}. We therefore adopt this potential for hydrogen, setting $Z=1$, and couple it to a linearly polarized laser pulse in the dipole approximation:
\begin{equation}
    V(x;t) = x \, \sqrt{I_0} f(t) \sin \omega t \quad ,
\end{equation}
where the envelope is given by $f(t) = \exp -((t-t_0)/\tau)^2$. The optical frequency of the pulse is taken as $\omega \approx 0.057 \, \text{au} = 780 \, \text{nm}$, lying in the near-infrared. The FWHM is $\tau \approx 4 \, \text{fs}$, such that centering the pulse at $t_0 \approx 10 \, \text{fs}$ covers it completely for $t \in [0,20] \, \text{fs}$. Intensities $I_0$ will vary between $1$ and $4 \times 10^{14} \, \text{W/cm}^2$. The initial state is chosen as the exact ground state for \eqref{eq:isc}, which we obtain numerically by diagonalizing the hamiltonian on a position grid ranging from $-1000$ to $1000 \, \text{au}$, with step-size $\Delta x \approx 0.1 \, \text{au}$. 

We note that the substitution of \eqref{eq:isc} by any other typical atomic potential, as well as performing simulations with more DoF, do not change the message contained in this manuscript: We are concerned only with the differences between classical and quantum, and dynamics due to LP pulses is essentially restricted to a single DoF \cite{majorosi2018improved}. Another important aspect is that quantum-classical agreement for systems that undergo tunneling is harder to achieve in one DoF than in more DoF. This was already remarked for a static electric field in \cite{spanner2003strong}, where the author correctly states that trajectory ``leakage'', which provides the trajectories necessary to reproduce ionization semiclassically, is more abundant in two DoF than in one. Thus, our choice of \eqref{eq:isc} as a model both simplifies interpretation and \emph{overestimates} the role played by TI. 

\noindent
\emph{Methods--} Quantum time-evolution is performed by solving the time-dependent Schr{\"o}dinger equation, where a split-operator method \cite{feit1982solution} using Blanes and Moan's 6th order algorithm is employed \cite{blanes2002practical}. The time-step chosen is $\Delta t=0.2 \, \text{au}$, corresponding to $1/550$ of an optical cycle. To mitigate wave reflections, we use a $\cos^{1/8}$ boundary mask placed at 10\% of the grid's extremities \cite{lorin2009mathematical, bezanson2017julia}.

Quantum dynamics in phase space, \emph{i.e.}~the time evolution of the Wigner function
\begin{equation}
    W_0(p_x,x) = \frac{1}{\pi \hbar} \int_\mathbb{R} \text{d}\gamma \, \langle x - \gamma \vert \psi_0 \rangle \langle \psi_0 \vert x + \gamma \rangle e^{2 i \gamma p/\hbar} \quad ,
\end{equation}
where $|\psi_0\rangle$ is the system's ground state, is dictated by Moyal's equation \cite{groenewold1946principles,moyal1949quantum,ozorio1998weyl,OzorioBook}. Here, unlike in other formulations, the $\hbar \to 0$ limit of time-evolution is well-defined and results in a von Neumann-like equation, with solution given by the Truncated Wigner Approximation (TWA)
\begin{equation}
    w(p_x,x;t) = (W_0 \circ \varrho_{-t}) (p_x,x) \quad . \label{eq:twa}
\end{equation}
In the above, $\varrho_{-t}$ is the hamiltonian flow bringing the point $(p_x,x)$ at $\tau=0$ to its value at $\tau=-t$. Computing this backwards flow requires us to employ a time-mirrored laser pulse in Hamilton's equations, which we solve using an adaptive-step algorithm with automatic stiffness detection \cite{DifferentialEquations.jl-2017,bezanson2017julia}. 

Expectation values using the TWA follow the recipe of the Wigner formalism: 
\begin{equation}
    \langle A(t) \rangle_{\rm{classical}} = \int_{\mathbb{R}^2} \text{d}x \, \text{d}p_x \, w(p_x,x;t) A(p_x,x) \quad , \label{eq:twa_means}
\end{equation}
where $A$ is the Weyl transform of operator $\hat{A}$ \cite{ozorio1998weyl}. We note that the usual way of computing classical expectation values is to transfer the time-dependence from $w$ to $A$, and then propagate forward in time \cite{mittal2020semiclassical}. We go through the trouble of negative times because, just as the marginals of Wigner functions describe quantum probability densities (PDs), the marginals obtained from the TWA can be seen as their classical equivalents. Comparisons between quantum and classical PDs will be our tool to determine whether or not TI is dominant: If this is the case, the classical PDs will have empty regions when compared to the quantum ones, indicating that the corresponding phase-space domain cannot be accessed by classical trajectories; If not, quantum and classical PDs will be non-zero on the same domain.  

\begin{figure}[t]
    \centering
    \includegraphics[width=0.5\textwidth]{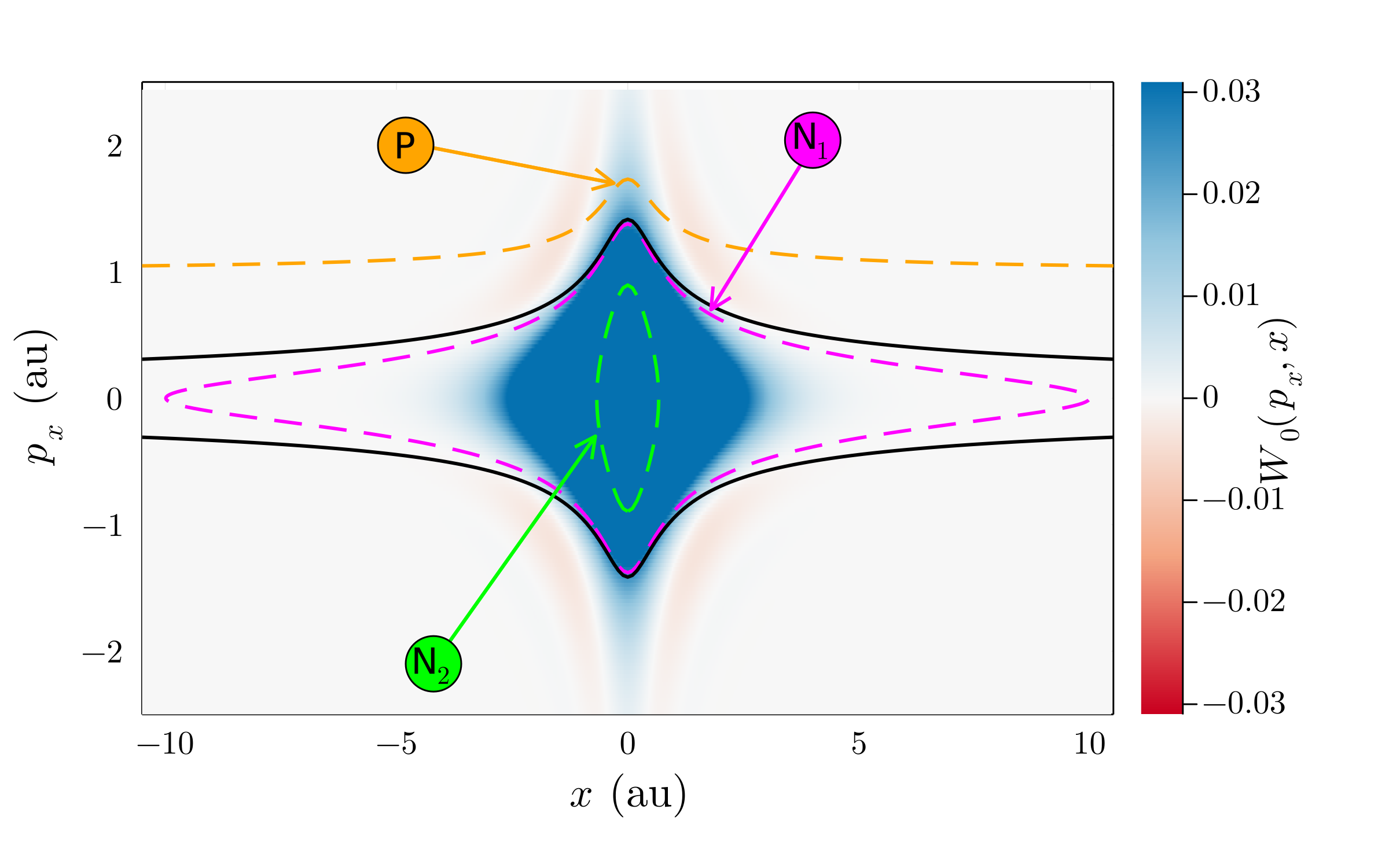}
    \caption{The ground state Wigner function, $W_0$. The solid black line is the zero energy contour, so orbits lying inside (outside of) it are bounded (scattered). The arrows mark orbits of type N$_1$ (magenta), N$_2$ (green) and P (orange).}
    \label{fig:ground_state}
\end{figure}

In practice, since \eqref{eq:twa_means} is performed by Monte Carlo, the classical (position) PDs are nothing but histograms of final positions in the TWA. These, in turn, come from a classically evolved, non-interacting gas of initial points usually chosen by importance sampling. Since we are interested in evolving the ground state Wigner function $W_0$, sampling the initial points according to $\vert W_0 \vert$ is very efficient, and we do so by employing a simple Metropolis-Hastings scheme \cite{Metropolis1953,Hastings1970}. The absolute value is necessary because the ground state of \eqref{eq:isc} is not a gaussian and, therefore, $W_0$ presents negative-valued regions (see Fig.~\ref{fig:ground_state} ahead) \cite{hudson1974wigner}. It should perhaps be mentioned that this association between a quantum state and a classical ensemble of points is a sensible way to establish quantum-classical correspondence for dynamics, being far more general than Ehrenfest's theorem \cite{ballentine1994inadequacy, drobny1997quantum, lasser2020computing, wang2021semiclassical}.

\noindent
\emph{Simulations--} In the case of potentials such as \eqref{eq:isc}, which is composed of bounded and scattering regions (corresponding to discrete and continuous quantum energy spectra), the initial points can have both closed and open trajectories. For \eqref{eq:isc}, closed and open trajectories have negative and positive energies, respectively, and we shall refer to them as being of type N or type P. 

Type N trajectories are all periodic. In the absence of the pulse some of them, which we call type N$_1$, naturally extend far beyond the mean ground state radius, orbiting the origin with large periods. Other type N trajectories, which we shall call N$_2$, remain near the origin and have small orbital periods. Since the laser pulse is in the near-infrared range, it acts on type N$_2$ and N$_1$ trajectories adiabatically and non-adiabatically, respectively. Thus, elementary classical perturbation theory \cite{Chirikov1979, ArnoldBook} tells us that the pulse will keep type N$_2$ trajectories mostly untouched, and scatter away some of type N$_1$. The former mechanism is responsible for the classical preservation of the ground state, since it conserves trajectories near the origin, and the latter is the one behind CAI. See Fig.~\ref{fig:ground_state} for a visual depiction of these concepts.

\begin{figure}[t]
    \centering
    \includegraphics[width=0.5\textwidth]{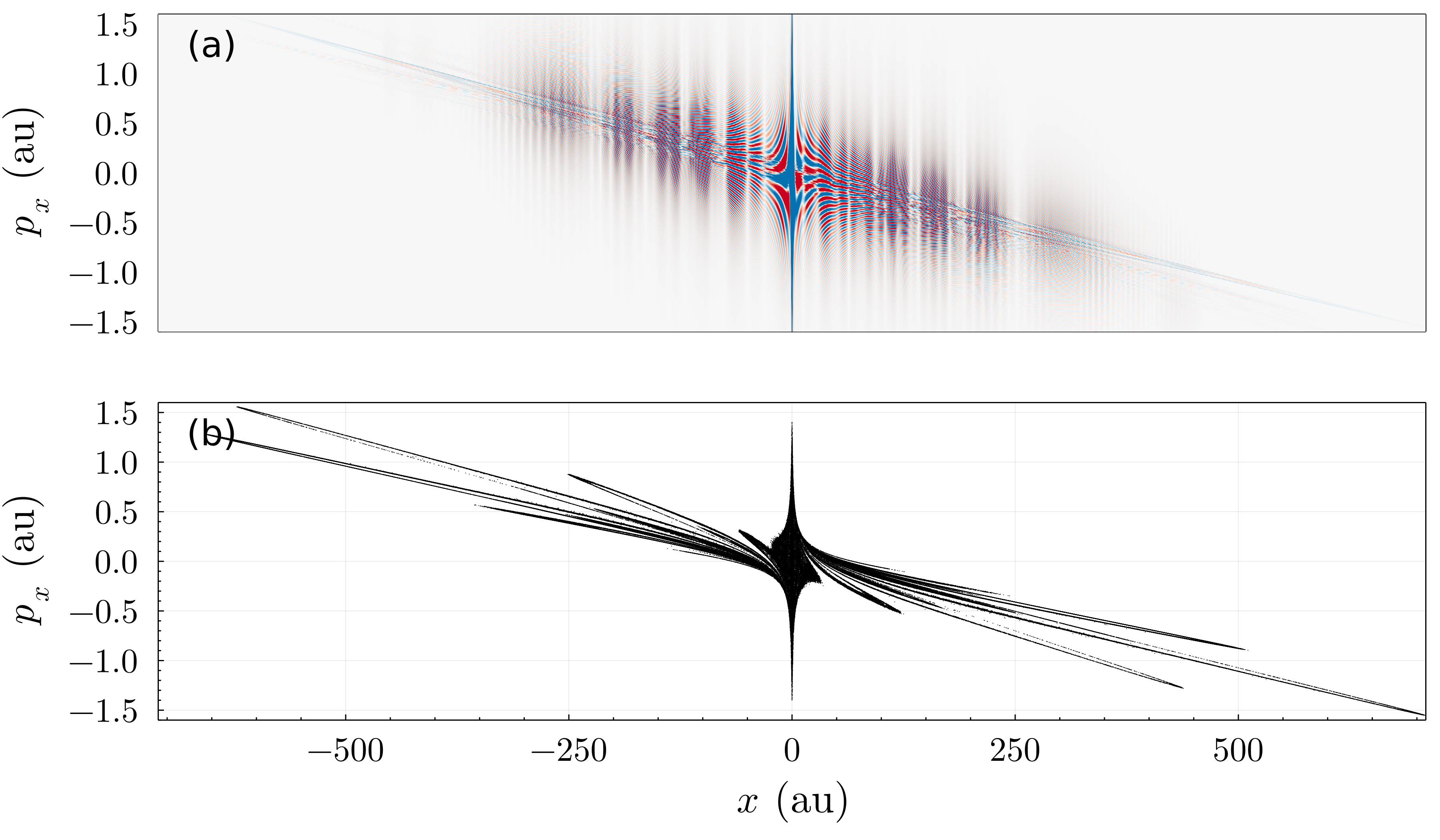}
    \caption{(a) Final Wigner function and (b) TWA after a pulse with intensity $I_0 = 2.0 \times 10^{14} \, \text{W/cm}^2$. Blue (red) colors in the Wigner function are positive (negative), and the TWA is built on around $10^6$ trajectories. The trajectories in panel (b) were all ionized by the laser field, since the ones lying on the tails of the Wigner function are absent in the set of initial points (see text). Note that the Wigner function and the TWA have essentially the same spread.}
    \label{fig:wigner_functions}
\end{figure}

The situation is very different with type P trajectories, whose initial points lie on the Wigner function's ``tails'' (see Fig.~\ref{fig:ground_state}). Due to the exponential decay of the tails, less than 0.01\% of the initial points form trajectories of type P. As these trajectories are not bound to the core, they are scattered away in a matter of a few attoseconds, even in the absence of the laser pulse. Moreover, by galilean invariance, we could simply translate the pulse to a time where type P trajectories would have already diverged, so there are many reasons to consider that they are irrelevant. After verifying that our results are unchanged whether or not we include them, we simply remove them. This filtering allows us to track classical ionization probabilities exclusively to the trajectories that \emph{truly} ionized, \emph{i.e.}~they started bounded and were scattered by the field. In Fig.~\ref{fig:wigner_functions} we show the Wigner function at the end of a pulse with intensity $I_0 = 2.0 \times 10^{14} \, \text{W/cm}^2$ together with the corresponding TWA, calculated using a filtered set of around $10^6$ initial points.

\begin{figure*}[t]
    \centering
    \includegraphics[width=\textwidth]{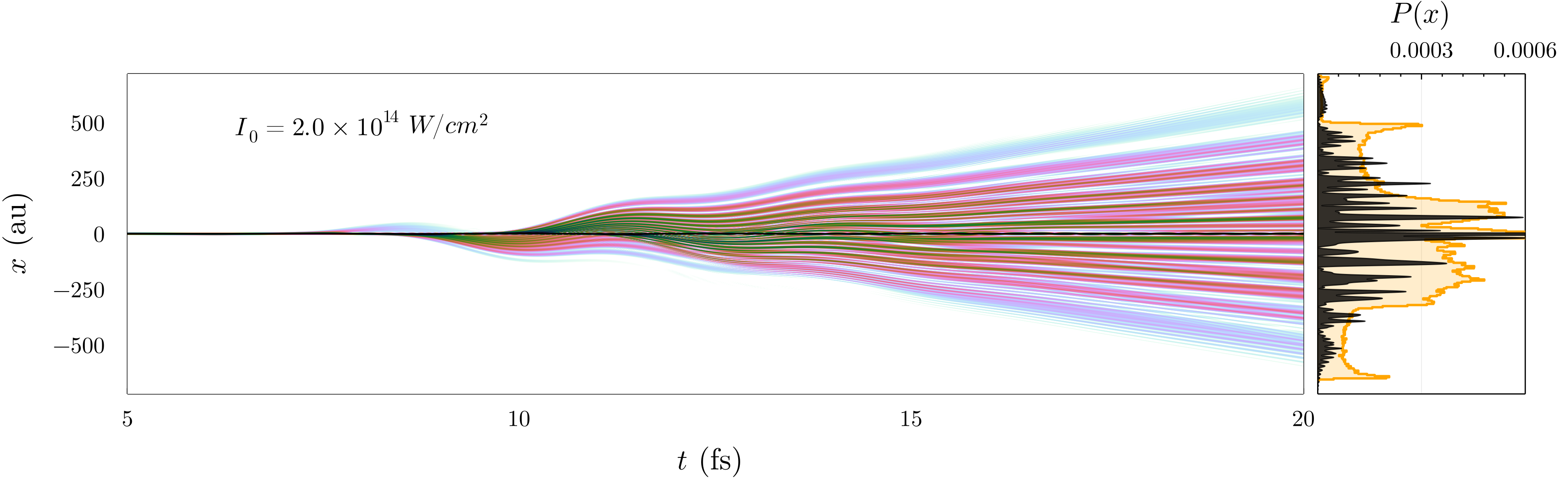}
    \caption{(Left) Heatmap of the probability of finding the electron as a function of space and time during the laser pulse. The color scale in the left panel is logarithmic, with green equal to $10^{-2}$ and purple to $10^{-5}$. (Right) Classical (orange) and quantum (black) position PDs at the end of the laser pulse. Note that the peak probability is close to $0.5$, but ionization probabilities are no larger than $0.0006$, requiring a massive zoom in order to be seen.}
    \label{fig:I0201e13}
\end{figure*}

Fig.~\ref{fig:wigner_functions} shows that CAI is taking place, since the TWA extends far beyond the core, and we now move on to quantify how much this classical spread accounts for total ionization. In the left panel of Fig.~\ref{fig:I0201e13} we display the quantum PD in the system as a function of time for the same intensity as Fig.~\ref{fig:wigner_functions}. Results are shown from $5 \, \text{fs}$ (which is when the laser pulse starts visibly acting on the state) to $20 \, \text{fs}$. In the right panel we show the final quantum PD, $P(x)$, together with its classical approximation, obtained from the TWA's position marginal. 

\noindent
\emph{Discussion--} In most situations where purely classical approximations are employed, what is expected is only a very rough sketch of what is obtained quantum mechanically. Classical physics cannot reproduce quantum superposition, which plays a fundamental role in several processes in strong-field physics (HHG, for instance, relies strongly on it \cite{van1999irregular}). This makes it impossible to reconstruct parts of PD that are strongly dependent on phase interference. The results of Fig.~\ref{fig:I0201e13} are surprising because the classical approximation is far more than a rough sketch: Not only does it correctly estimate the (minuscule) order of atomic ionization probability, but it also highlights the overall structure of the quantum result. Most importantly, the classical PD extends over essentially the same range as the quantum one despite a Keldysh parameter smaller than unity, namely $\gamma_\text{K} \approx 0.78$, which should be indicative of a strong presence of tunneling \cite{GrossmannBook}.

The sharp peaks displayed by the classical PD in the right panel of Fig.~\ref{fig:I0201e13} are due to the ``whorls'' and ``tendrils'' of Fig.~\ref{fig:wigner_functions}, \emph{i.e.}~the initial state is deformed into filaments that are sheared and folded, and the classical position marginals have peaks on the filaments lying perpendicularly to the momentum axis \cite{berry1979evolution, berry1979quantum}. This filamentary structure is usually not enough to reproduce peak heights, and one must resort to semiclassical approximations: Quantum interference is then achieved by a rigorous endowing of classical trajectories with accumulated phases, which superpose and correct peak intensities \cite{Maslov1981,lando2019quantum}. What is fundamental to our purposes is that only the extremities of the PD in Fig.~\ref{fig:I0201e13}, which are barely visible, are not classically accessible -- These are the ones that emerge exclusively from TI.

\begin{figure}[t]
    \centering
    \includegraphics[width=0.5\textwidth]{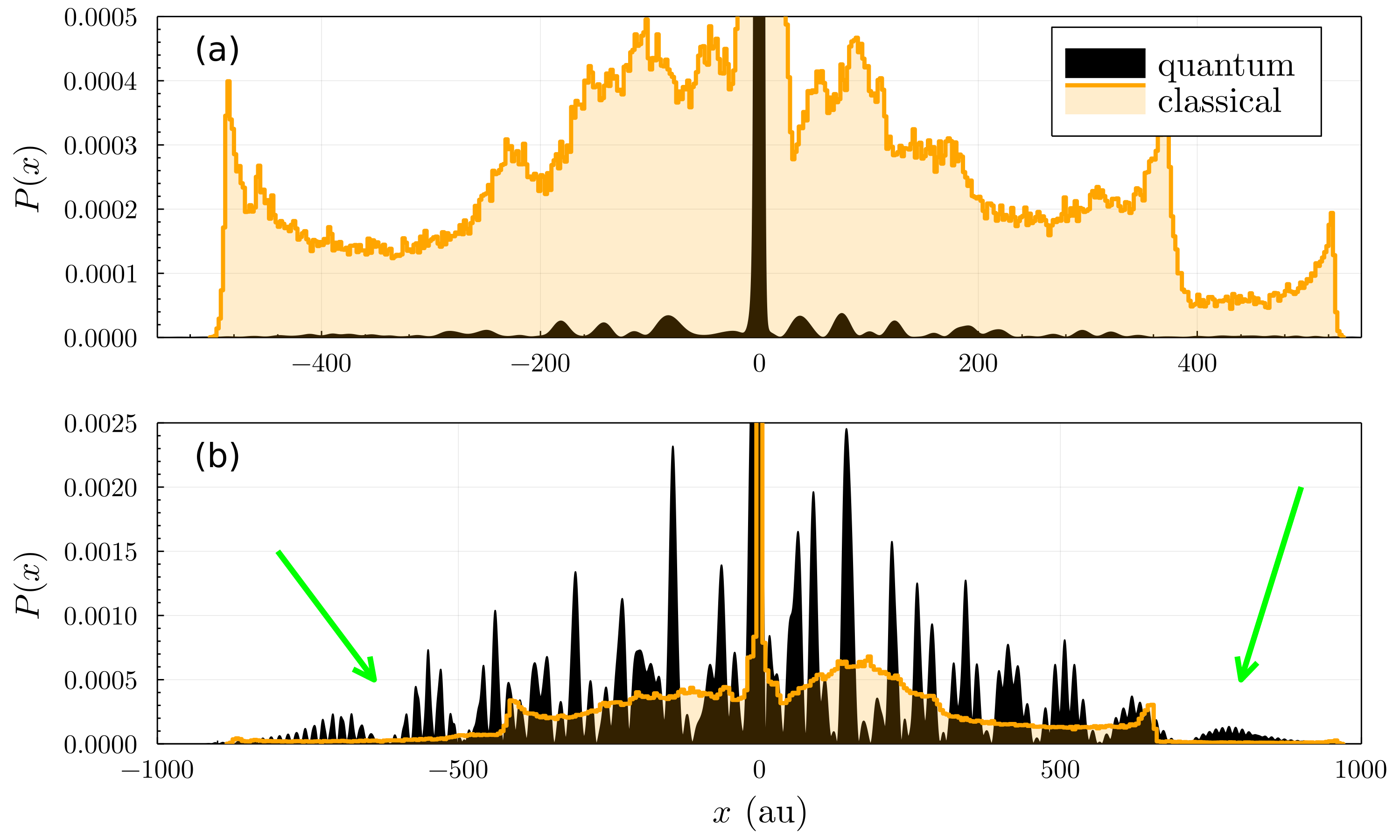}
    \caption{Quantum (black) and classical (orange) PDs at the end of the laser pulse for: (a) $I_0 = 1.0 \times 10^{14} \, \text{W/cm}^2$; (b) $I_0 = 4.0 \times 10^{14} \, \text{W/cm}^2$. Classical calculations employed $10^6$ trajectories. The arrows in panel (b) point at regions where TI can be safely disentangled from CAI.}
    \label{fig:hists}
\end{figure}

The success of both classical and semiclassical mechanics, however, depends on the existence of available classical trajectories. After they are all scattered away, TI becomes the only available escape pathway from the core, as can be seen for static fields or long laser pulses \cite{spanner2003strong}. If the pulses are strong, trajectories will be scattered faster, such that very strong fields should also pose problems for classical/semiclassical propagation \footnote{Note that this is in line with standard Keldysh theory, but for different reasons.}. Thus, in Fig.~\ref{fig:hists} we display the final probability densities for two different laser pulse intensities: One weaker ($\gamma_\text{K} \approx 1.1$) and the other one stronger ($\gamma_\text{K} \approx 0.54$) than in Figs.~\ref{fig:wigner_functions} and \ref{fig:I0201e13}. 

As we can see in Fig.~\ref{fig:hists}(a), for ``weak'' fields the quantum PD is completely supported by the classical one, which is almost one order of magnitude larger. This shows that, in the limit of weak fields, CAI is notoriously dominant over TI, so much so that it even resembles a case of strong localization \cite{anderson1958absence}. Whether or not semiclassical corrections will be able to suppress the classical PD and reproduce the peaks depends on how far one is in the semiclassical regime, \emph{i.e.}~how large the actions associated to the trajectories are with respect to $\hbar$, but a perfect reproduction of probability densities is not the objective of this manuscript. What is important is that there's no region in position space that classical trajectories didn't reach, so TI cannot be dominant. In Fig.~\ref{fig:hists}(b) however, we note the increasing presence of peaks falling outside the reach of classical trajectories, and TI is seen to arise from outside in. Using Fig.~\ref{fig:I0201e13}, we can describe this quite easily: Some portions of the wave function that ionize near the maxima of the laser pulse, with the absolute maximum at $t_0 = 10 \, \text{fs}$, go farther. They provide peaks at the PD's extremities that cannot be fully reached by CAI, although this is hard to see for the intensity used in Fig.~\ref{fig:I0201e13} and much clearer in Fig.~\ref{fig:hists}(b). In addition to the classically unreachable extremities, it is also unlikely that semiclassical approximations will be able to resolve most of the peaks in Fig.~\ref{fig:hists}(b), as it is clear that both TI and CAI are taking place. 

Interestingly, if we interpret Fig.~\ref{fig:hists}(b) together with Fig.~\ref{fig:I0201e13}, we see that the purest tunneling contributions can be tracked to portions that ionize and never recombine -- In the terminology of HHG literature, the classical trajectories near these portions (if there are any) never \emph{recollide} with the core \cite{protopapas1996recollisions,GrossmannBook,dubois2020envelope}. The recolliding trajectories abide, as expected, within a ball of quiver radius $\alpha_0 = \sqrt{I_0}/\omega^2$ \cite{GrossmannBook}. As can be seen in the Figs.~\ref{fig:wigner_functions}, \ref{fig:I0201e13} and \ref{fig:hists}, however, this region is always well-populated by classical trajectories. This explains why HHG spectra could be obtained in \cite{zagoya2014quantum} through semiclassical approximations: Tunneling contributions might even be present near the core, but the classical trajectories in the region were enough to reproduce spectra exclusively from CAI. 

It is then reasonable that the first step in semiclassical step models be called ``ionization'' instead of ``tunneling ionization'', since one cannot effectively know whether it came from tunneling or not. This, moreover, adds to the previously raised concerns about the difficulty of theoretically and experimentally resolving tunneling time ambiguities, since the contributions from TI and CAI might be impossible to disentangle. In the end, it is possible that the measured angular delays in attoclocks, often interpreted as the ``time spent inside a barrier'', are predominantly due to CAI and directly traceable to transport properties of classical trajectories. This would render ``tunneling times'' essentially unrelated to tunneling. 

\noindent
\emph{Conclusions--} We have used a one-dimensional atom to bring forward the fact that isolating tunneling from classically allowed ionization can be extremely hard, even for the simplest of systems, when they are interacting with intense and short laser pulses. Our results show that what is known as ``tunneling ionization'' in semiclassical step models in atomic physics has a major classical fingerprint, with direct consequences to the proper interpretation of higher-harmonics generation and tunneling times, among others. \\

\noindent
\emph{Acknowledgements--} I thank Alfredo Ozorio de Almeida, Andrew Hunter, Denis Ullmo, Frank Gro{\ss}mann, Jessica Almeida, Jonathan Dubois, Olivier Giraud, Peter Schlagheck, Sebastian Gemsheim and Steven Tomsovic for many stimulating discussions. I also thank Jan-Michael Rost and the hospitality of the Max Planck Institute for the Physics of Complex Systems, where the initial stages of this work were carried out.

\bibliography{Bibliography.bib}

\end{document}